\documentclass[aps,prl,twocolumn,showpacs,floatfix,superscriptaddress]{revtex4}

\usepackage{graphicx}
\usepackage{epsfig}
\usepackage{amsmath}    
\usepackage{amssymb}    

\begin{document}
 

\title{Pairwise Network Information and Nonlinear Correlations}

\author{Elliot A. Martin}
\affiliation{Complexity Science Group, Department of Physics and
  Astronomy, University of Calgary, Calgary, Alberta, Canada, T2N 1N4}
\author{Jaroslav  Hlinka}
\affiliation{Institute of Computer Science, The Czech Academy of Sciences, Pod vodarenskou vezi 2, 18207 Prague, Czech
  Republic}
\affiliation{National Institute of Mental Health, Topolov\'{a} 748, 250 67 Klecany, Czech Republic}
\author{J\"{o}rn Davidsen} 
\email[]{davidsen@phas.ucalgary.ca}
\affiliation{Complexity Science Group, Department of Physics and
  Astronomy, University of Calgary, Calgary, Alberta, Canada, T2N 1N4}

\date{\today}

\begin{abstract}  
  Reconstructing the structural connectivity between interacting units
  from observed activity is a challenge across many different
  disciplines. The fundamental first step is to establish whether or
  to what extent the interactions between the units can be considered
  pairwise and, thus, can be modeled as an interaction network with
  simple links corresponding to pairwise interactions.  In principle
  this can be determined by comparing the maximum entropy given the
  bivariate probability distributions to the true joint entropy. In
  many practical cases this is not an option since the bivariate
  distributions needed may not be reliably estimated, or the
  optimization is too computationally expensive.  Here we present an
  approach that allows one to use mutual informations as a proxy for
  the bivariate probability distributions. This has the
  advantage of being less computationally expensive and easier to
  estimate.  We achieve this by introducing a novel entropy
  maximization scheme that is based on conditioning on entropies and
  mutual informations. This renders our approach typically superior to
  other methods based on linear approximations. The advantages of the
  proposed method are documented using oscillator networks and a
  resting-state human brain network as generic relevant examples.
\end{abstract}


\maketitle


Pairwise measures of dependence such as cross-correlations (as
measured by the Pearson correlation coefficient or covariance matrix)
and mutual information are widely used to characterize the
interactions within complex systems. They are a key ingredient to
techniques such as principal component analysis, empirical orthogonal
functions, and functional networks (networks inferred from dynamical
time series)~\cite{Donges2015::CD,haimovici13,timme14}. These
techniques are widespread since they provide greatly simplified
descriptions of complex systems, and allow for the analysis of what
might otherwise be intractable problems~\cite{bullmore09}. In
particular, functional networks have been widely applied in fields
such as neuroscience~\cite{bullmore09,Eguiluz2005::PRL},
genetics~\cite{Margolin2006::BMCbioinfo}, and cell
physiology~\cite{Stovzer2013::PLoSComBio}, as well as in climate
research~\cite{Donges2015::CD,Runge2015::NC}.

In this paper we study how faithfully these measures alone can
represent a given system. With the increasing use of functional
networks this topic has received much attention recently, and many
technical concerns have been brought to light dealing with the
inference of these networks. Previous studies have shown that the
estimates of the functional networks can be negatively affected by
properties of the time
series~\cite{Martin2013::EPL,Palu2011::NPG,bialonski11}, as well as
properties of the measure of association,
e.g. cross-correlations~\cite{Tirabassi2015::SciRep,Hlinka2012::Chaos,martin2014::NPG,mader15}. In
this work however, we address a more fundamental question: How well do
pairwise measurements represent a system?

In principle this can be evaluated using a maximum entropy approach. The corresponding
framework was first laid out in~\cite{Schneidman2003::PRL} and later applied
in~\cite{Schneidman2006::Nat}, where they assessed the rationale of
only looking at the pairwise correlations between neurons.  They
examined how well the maximum entropy distribution, consistent with
all the pairwise correlations described the system. If the
system is not well described by this maximum entropy distribution then
we know from the work of Jaynes~\cite{Jaynes1957::PRa} that other
information beyond pairwise relationships would need to be taken into
account. Similar analyses have since been applied in
neuroscience~\cite{Watanabe2013::NatCom,Yu2011::JNeurosci,ohiorhenuan10},
as well as in genetics~\cite{Lezon2006::PNAS},
linguistics~\cite{stephens10}, economics~\cite{Xi2014::PhysicaA}, and
to the supreme court of the United States~\cite{Lee2013::JStatPhys}.

However, the data to accurately estimate the needed bivariate
probability distributions may not be available. To get around this
some researchers have used the first two moments of the variables as
constraints instead of the full bivariate
distributions~\cite{Bialek2012::PNAS,Wood2012::PNAS} --- effectively
using the cross-correlations as their constraints. In the case of
binary variables, as in the original work~\cite{Schneidman2006::Nat},
this is equivalent to conditioning on the bivariate distributions. For
larger cardinality variables this is only an approximation though, as
the cross-correlation is only sensitive to linear
relationships~\cite{kantz}. Systems where larger cardinalities and
nonlinear behaviour are thought to play a significant role such as in
coupled oscillators --- which have been used to model systems as
diverse as pacemaker cells and crickets~\cite{Pikovsky2003} --- are,
however, rather the norm than an exception~\cite{kantz}. In
particular, we show here that this plays a significant role in a
resting-state human brain network.

In order to retain the attractive properties of the cross-correlation
and simultaneously capture a much wider range of relationships we
propose using the mutual information. Mutual information can detect
arbitrary pairwise relationships between variables, and is only
non-zero when the variables are pairwise independent, making it the
ideal measure~\cite{cover}. However, while calculating the maximum
entropy given the moments of a distribution results in simple
equations in the probabilities, using mutual informations as
constraints results in transcendental equations which are much harder
to solve.  We circumvent this problem here using the set theoretic
formulation of information theory~\cite{Yeung2008information}, which
gives us an upper bound on the maximum entropy that is saturated in
many cases.

\begin{figure}[!h]
  \begin{center}
    \includegraphics*[width=0.6\columnwidth]{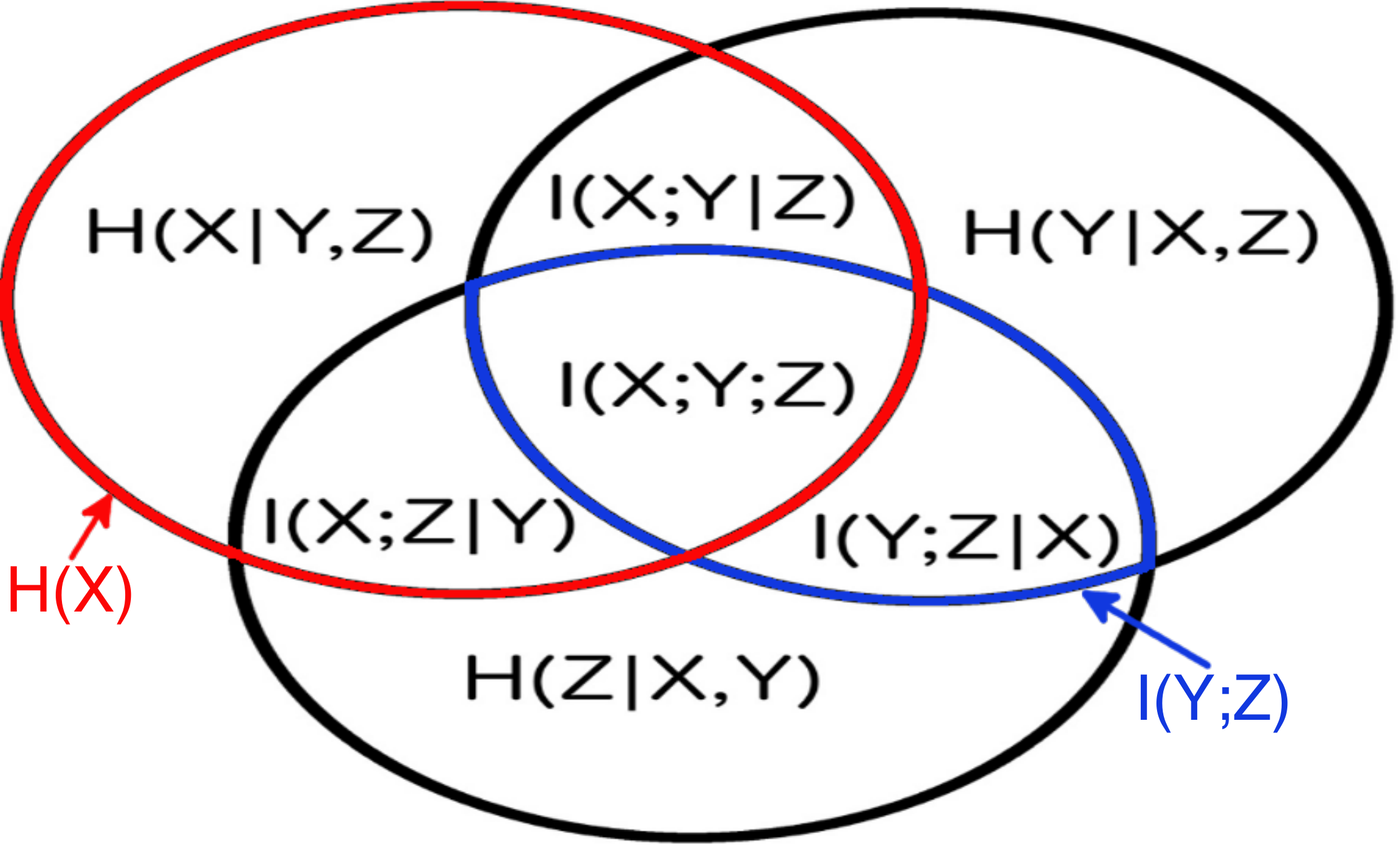}
   \end{center}
   \caption{\label{Fig:InfoDiag} (Color online) The information
     diagram for 3 variables. It contains 7 regions corresponding to
     the possible combinations of 3 variables, with their
     corresponding information theoretic quantities defined in the
     text. The univariate entropy $H(X)$ is the sum of all the regions
     in the red circle, and the mutual information $I(Y;Z)$ is the sum
     of all the regions in the blue oval.}
\end{figure}

The set theoretic formulation of information theory allows us to map
information theoretic quantities to the regions of an information
diagram, a variation of a Venn diagram. The information diagram for
three variables is shown in Fig.~\ref{Fig:InfoDiag} with the
associated information theoretic quantities labeled
\footnote{We use the convention $p(x,y,z) = P(X = x, Y = y, Z = z)$.}: 
entropy, $H(X) = -\sum p(x) \log(p(x))$; conditional entropy,
$H(X|Y,Z) = -\sum p(x,y,z) \log(p(x | y, z))$; mutual information,
$I(X;Y) = \sum p(x,y) \log(p(x,y)/(p(x)p(y)))$; conditional mutual
information, $I(X;Y|Z) = \sum p(x,y,z) \log
\left(p(x;y|z)/[p(x|z)p(y|z) ] \right)$; multivariate mutual
information, $I(X;Y;Z) = I(X;Y) - I(X;Y|Z)$.

We illustrate our method using systems of coupled oscillators, as they
commonly occur in nature and are used to model a large variety of
systems~\cite{Pikovsky2003}.  In particular we look at the Kuramoto
model~\cite{Kuramoto1975,Acebron2005::RMP} as a paradigmatic example
that is capable of a wide range of dynamics from synchronization to
chaos~\cite{Maistrenko2004::PRL}, and hence provides an excellent test
bed for our method.

\paragraph{Method:}

Given a set of $N$ variables ($\{X\}_N$), we want to know how well the
cross-correlation or mutual information between all pairs of variables
can encode the state of the system. To do this we first determine the
maximum entropy consistent with the given measure of similarity,
$H_m(\{X\}_N)$, which represents a standard variational problem.
This means that any model of the system consistent with the ${N
  \choose 2}$ values of the similarity measure can have an entropy of
at most $H_m(\{X\}_N)$. From the work of
Jaynes~\cite{Jaynes1957::PRa}, we also know that any model of the
system with a smaller entropy must implicitly or explicitly include
information beyond these values. As a result the true joint entropy,
$H(\{X\}_N)$, will always be less than or equal to $H_m(\{X\}_N)$.

If the variables are all independent the entropy of the system is the
sum of the entropies of the individual variables, $H_I(\{X\}_N) =
\sum_i H(X_i)$. The most this uncertainty can be reduced is if the
true joint entropy is known, and is given by the multi-information
(also called total correlation~\cite{Watanabe1960::IBM}),
$I_N(\{X\}_N) = H_I(\{X\}_N) - H(\{X\}_N) \geq 0$. We similarly define
the measure information $I^m(\{X\}_N) = H_I(\{X\}_N) - H_m(\{X\}_N)$,
to be the reduction in uncertainty given a measure. The fraction of
information retained by describing the system with a given measure, as
opposed to the true joint entropy, is then $0 \leq I^m/I_N \leq 1$. If
the used measure is the bivariate probability distribution, we call
$I^m$ the pairwise network information, or the second order connected
information as defined in~\cite{Schneidman2003::PRL}. This is
approximated linearly if the measure used is the cross-correlation,
and nonlinearly if the measure used is the mutual information.

When using the cross-correlation, estimating $H_m$ is conceptually
straightforward, though finding the optimum value can be
computationally expensive. Estimates of the first two moments of the
variables uniquely determine the cross-correlations, and can be used
as constraints in a Lagrange multiplier problem solving for $H_m$. The
resulting probability distribution $P_m({\{X\}_N})$ is the Boltzmann
distribution $ P_m(\{X\}_N) = \exp{ \left( \sum_i h_i x_i + \sum_{i
      \geq j}J_{i,j}x_ix_j \right)},$ where $h_i$ and $J_{i,j}$ are
the Lagrange multipliers~\cite{timme14}.

When using mutual information, estimating $H_m$ with Lagrange
multipliers is much harder as the derivatives of the Lagrange function
are transcendental functions in $P_m(\{X\}_N)$. Instead, we use the
mutual informations and univariate entropies as constraints, and draw
on the structure of information diagrams. Each univariate entropy and
mutual information corresponds to a region in the information diagram
that can be written as a sum of a number of {\it atomic regions
  (atoms)}. The sum over all atoms is simply $H(\{X\}_N)$.  Thus, as
seen in Fig.~\ref{Fig:InfoDiag}, we obtain constraints of the form:
\begin{align}
\label{Eq:IConst}
\text{const} = I(Y;Z) =&  I(Y;Z|X) + I(X;Y;Z),\\
\nonumber
\text{const} =  H(X) =&  H(X|Y,Z) + I(X;Y|Z) \\
&+ I(X;Z|Y) + I(X;Y;Z).
\end{align}
In general, a system of N variables results in ${N \choose 1}$
univariate entropy constraints, ${N \choose 2}$ mutual information
constraints, and $A = \sum_{k=1}^N {N \choose k} = 2^N -1$ atoms to be
determined. In the simplest case of $N = 3$ variables we have six
constraints and $A = 7$ regions to specify, see
Fig~\ref{Fig:InfoDiag}. This means we only have one free parameter,
making the maximization process to get $H_m(\{X\}_N)$ particularly
easy in this case; in general there are $\sum_{k=3}^N {N \choose k}$
free parameters.


Apart from the chosen constraints defined above, there are also
general constraints on the values of the subregions ensuring they
define a valid information diagram, i.e. that there exists a
probability distribution with corresponding information-theoretic
quantities.  A family of such constraints (so-called Shannon
inequalities) can be inferred from the fundamental requirement that,
for discrete variables, (conditional) entropies and mutual
informations are necessarily non-negative: A) $H(X_i | \{X\}_N - X_i)
\geq 0$; B) $I(X_i;X_j| \{X\}_K) \geq 0$, where $i \neq j$ and
$\{X\}_K \subseteq \{X\}_N - \{X_i,X_j\}$
\footnote{This set of equalities is minimal in the sense that no
inequality is implied by any combination of the
others~\cite{Yeung2008information}.}.
Each inequality can also be
written as a sum of atoms, e.g. 

\begin{equation}
  \label{Eq:ShannonIneq}
  I(X_1;X_2|X_3) = I(X_1;X_2|X_3,X_4) + I(X_1;X_2;X_4|X_3) \geq 0.
\end{equation}


Not so-well known, for $N\geq 4$, there are also inequalities that are
not deducible from the Shannon inequalities, so called non-Shannon
inequalities~\cite{Yeung2008information}.  In principle, these
inequalities may be included in our maximization problem; however,
they have not yet been fully described. Therefore, we suggest
constructing the diagram with the maximum entropy that satisfies the
problem specific constraints and is consistent with the Shannon
inequalities.  As it may violate the non-Shannon inequalities, it may
not represent a valid distribution. However, the sum of the atomic
regions would still be an upper bound on the entropy $H_m$, and thus
provide a \emph{lower} bound on $I^m/I_N$.  Notably, for the
particular (and in our simulations common) case where all $A$ elements
are non-negative --- which is always true for $N=3$ --- one can prove
that the bound is attainable (see Theorem
3.11~\cite{Yeung2008information}).


To summarize, the task of finding the maximum entropy conditioned on
the univariate entropies, mutual informations, and elemental Shannon
inequalities, can be solved using linear optimization: Each constraint
will take the form of a linear (in-)equality, as in
Eqs.~\eqref{Eq:IConst},~\eqref{Eq:ShannonIneq}, and we maximize the
N-variate entropy by maximizing the sum over all $A$ atoms of the
information diagram. Thus, we avoid having to perform the maximization
over probability distributions.

\paragraph{Example of an Nonlinear Pairwise Distribution:}

We now give an example illustrating how the mutual information can
better detect pairwise relationships then the
cross-correlation. Consider a set of variables $\{X\}_N$: each
variable is drawn uniformly from the set $\{-1,0,1\}$, and all
variables are simultaneously $0$ or independently distributed among
$\{-1,1\}$. The cross-correlation between any pair of variables is
zero, and therefore consistent with the hypothesis that all variables
are independent. Therefore, the fraction of information captured by
the cross-correlation is $I^m/I_N = 0$. However, there is a
significant amount of mutual information between the variables.

Since $P(X_i|\{X\}_N - X_i) = P(X_i|X_{j\neq i}) \forall i$ and $j$,
all the conditional mutual informations are zero. Therefore, the only
nonzero atoms in the information diagram will be the N-variate mutual
information $I(X_1;...;X_N) = I(X_1;X_2)$, and the conditional
entropies $H(X_i| \{X\}_N - X_i) = 2/3$ bits. This is the maximum
entropy diagram consistent with the pairwise mutual informations and
univariate entropies, so the expected result using the mutual
information is $I^m/I_N = 1$.  We can see why this is the case by
starting with the information diagram for 2 variables (which is fixed
from our conditions), and successively adding new variables. The
addition of each new variable adds $2/3$ bits to the total entropy ---
which is the maximal amount consistent with the mutual informations.

\paragraph{Kuramoto Model:}

The Kuramoto model is a dynamical system of $N$ phase oscillators with
all to all coupling proportional to
$K$~\cite{Kuramoto1975,Acebron2005::RMP}. The $ith$ oscillator has an
intrinsic frequency of $\omega_i$, a phase of $\theta_i$, and its
dynamics is given by
$  \frac{\partial \theta_i }{\partial t} = \omega_i + \frac{K}{N} \sum_{j=1}^N\sin(\theta_j - \theta_i) + \eta_i(t).
$
Here, we have followed~\cite{Sakaguchi1988::PTP} and added a dynamical
noise term to mimic natural fluctuations and environmental effects;
$\eta_i(t)$ is drawn from a Gaussian distribution with correlation
function $\langle \eta_i(t)\eta_j(t') \rangle = G
\delta_{i,j}\delta(t-t')$, where $G$ determines the amplitude of the
noise.
For values of $K$ above a critical threshold, $K > K_{c}$,
synchronization occurs~\cite{Pikovsky2003}. In the limit of constant
phase differences the dynamics are trivial, and knowledge of one
oscillator will specify the phase of all others. Therefore, pairwise
information is sufficient to describe the system in this case. Yet,
the presence of noise results in random perturbations of the phases
and typically prevents constant phase
differences~\cite{Sakaguchi1988::PTP} such that only $I^m/I_N \lesssim
1$ is expected. In the weak coupling regime when synchronization is
absent, it is nontrivial what $I^m/I_N$ should be.

To estimate $I^m/I_N$ and to establish the importance of the level of
discretization or cardinality, we first discretize the phase of each
oscillator into $n$ equally likely states \footnote{Using $n$ equally
  sized bins did not noticeably alter our results.}. Alternatively,
estimators for continuous variables can be used as we discuss
in~\cite{Martin2015::PRE}.
To provide clear proofs of principle, we first focus on three-oscillator
systems in the following as this is the smallest system size at which
the results are non-trivial. Specifically, we consider three different
cases: (i) all oscillators have the same intrinsic frequency, (ii) all
oscillators have unique intrinsic frequencies and are still
synchronized, and (iii) all oscillators have unique intrinsic
frequencies and the entire system and all subsystems are
unsynchronized (``weak coupling regime''). For three-oscillator
systems, the corresponding parameter regimes in the absence of noise
have been carefully documented in Ref.~\cite{Maistrenko2004::PRL}.

For each of the three cases examined we created ensembles of
100 three-oscillator systems, where each element of the ensemble will
have randomly sampled frequencies
\footnote{For the numerical simulations, we use the Euler-Maruyama
  method, with a time step $dt = 2^{-6}$, and unless otherwise stated
  we use an integration time of $T=2000$ in all of our results. We
  also discard times up to $50$ to remove transient effects, and
  resample the data only taking every $8th$ data point.}.
These ensembles are studied in two different noise regimes, $G =
0.001$ and $G = 0.5$. The same ensemble of frequencies is used in both
noise regimes.

\begin{figure}[!h]
  \begin{center}
    \includegraphics*[width=0.99\columnwidth]{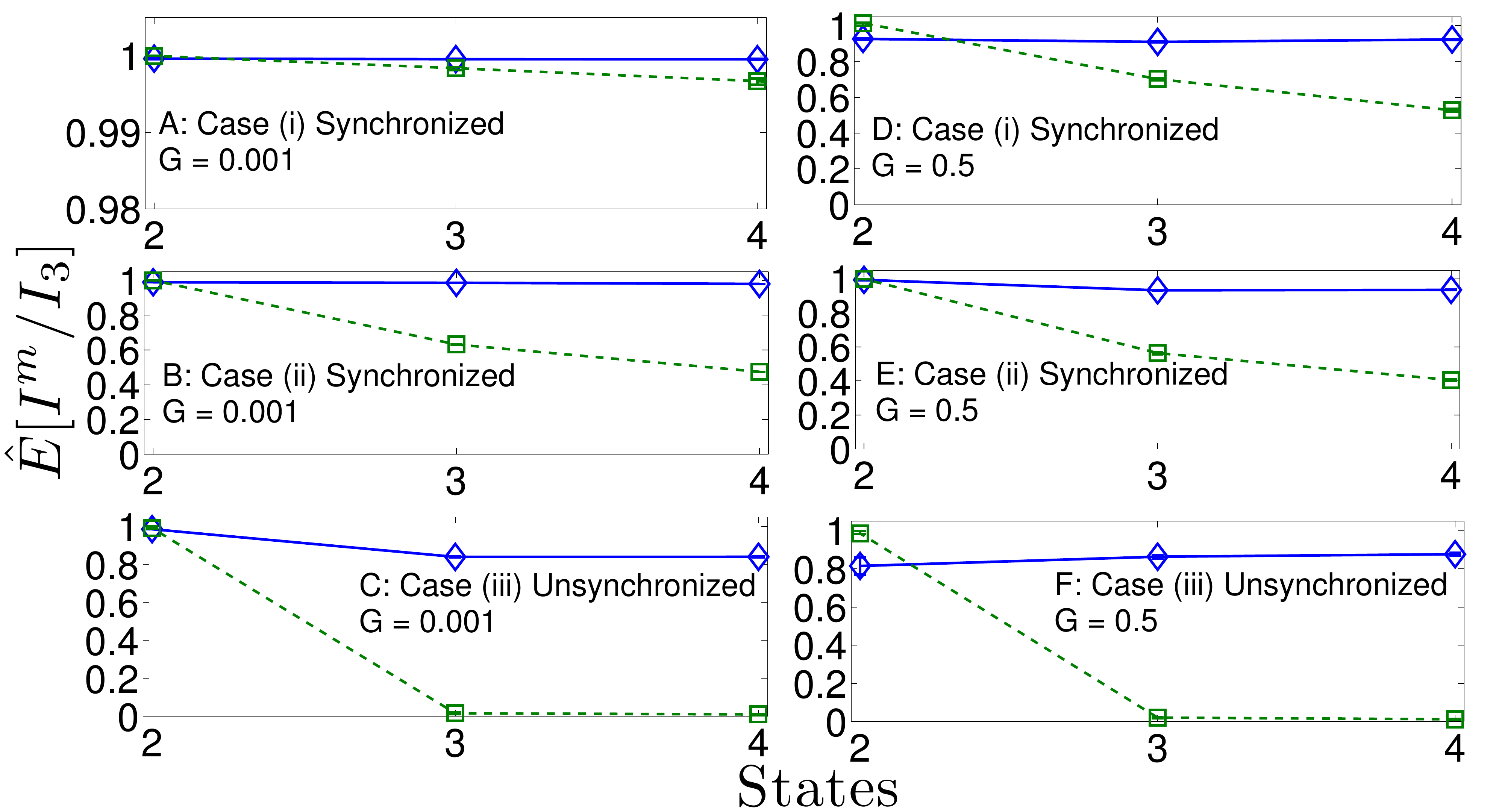}
   \end{center}
   \caption{\label{Fig:Completeness} (Color online) The fraction of
     shared information coded by the mutual information (blue
     diamonds, solid line) and the cross-correlation (green squares,
     dashed line). Notice the scale only goes from $0.98-1$ in
       Panel A, and from $0-1$ for the rest. The estimated
     expectations, $\hat{E}[...]$, are averages over the ensemble of
     100 realizations where we draw $\omega_3$ from a normal
     distribution with zero mean and unit variance. Uncertainties
     corresponding to the $25\%$ and $75\%$ quantiles are smaller than
     the symbol sizes.}
\end{figure}

In the first case all oscillators are synchronized with $\omega_1 =
\omega_2 = \omega_3$, and $K = 1.65$. Recall, in the synchronized case
we expect $E[I^m/I_N] \approx 1$. This is indeed what we see in the
low noise case, $G = 0.001$, Fig.~\ref{Fig:Completeness}~A; though the
mutual information preserves slightly more information at larger
cardinalities. However, for increased noise, $G=0.5$, the
cross-correlation performs poorly at larger cardinalities, while the
mutual information behaves robustly, Fig.~\ref{Fig:Completeness}~D.

For the second case, where the oscillators are synchronized with
different intrinsic frequencies, we use $\Delta_1/ \Delta_2 = 1.11$,
$K/ \Delta_2 = 4$, and $K = 2.20$, where $\Delta_1 = \omega_2 -
\omega_1$, and $\Delta_2 = \omega_3 - \omega_2$. Now at both noise
levels, at cardinalities greater than 2, the cross-correlation fails
to capture a significant portion of the available information --- as
$\hat{E}[I^m/I_N]$ is significantly less than one ---
Fig.~\ref{Fig:Completeness}~B and E. This indicates that even small
amplitude noise can prevent the cross-correlation from accurately
encoding information about the system in this case.  The mutual
information again robustly encodes almost all of the possible
information, $\hat{E}[I^m/I_N] \approx 1$, in both noise regimes and
across all discretizations analyzed.

In the final case, the weak coupling regime ($K/ \Delta_2 = 0.99$, all
other parameters as in the second case), we do not have a strong
hypothesis for what $E[I^m/I_N]$ should be. In
Fig.~\ref{Fig:Completeness}~C and F we can see that the
cross-correlation encodes virtually no information about the system
for cardinalities greater than 2, $\hat{E}[I^m/I_3] \approx 0$.  The
mutual information again robustly encodes the vast majority of the
multi-information, with $\hat{E}[I^m/I_3] > 0.8$ for all noise levels
and discretizations.

Similar overall results hold for larger systems and when only a subset of
oscillators is observed. As an example, we consider here a system of 100 non-identical Kuramoto
oscillators in two regimes: i) All oscillators are synchronized,
$K=4$; ii) the oscillators are partially synchronized with more than
20 different synchronized clusters, $K=1.75$. In both cases we use the
same set of intrinsic frequencies (drawn from a normal distribution
with mean zero and unit variance), and a noise level $G = 0.001$.

\begin{figure}[!h]
  \begin{center}                         
    \includegraphics*[width=\columnwidth]{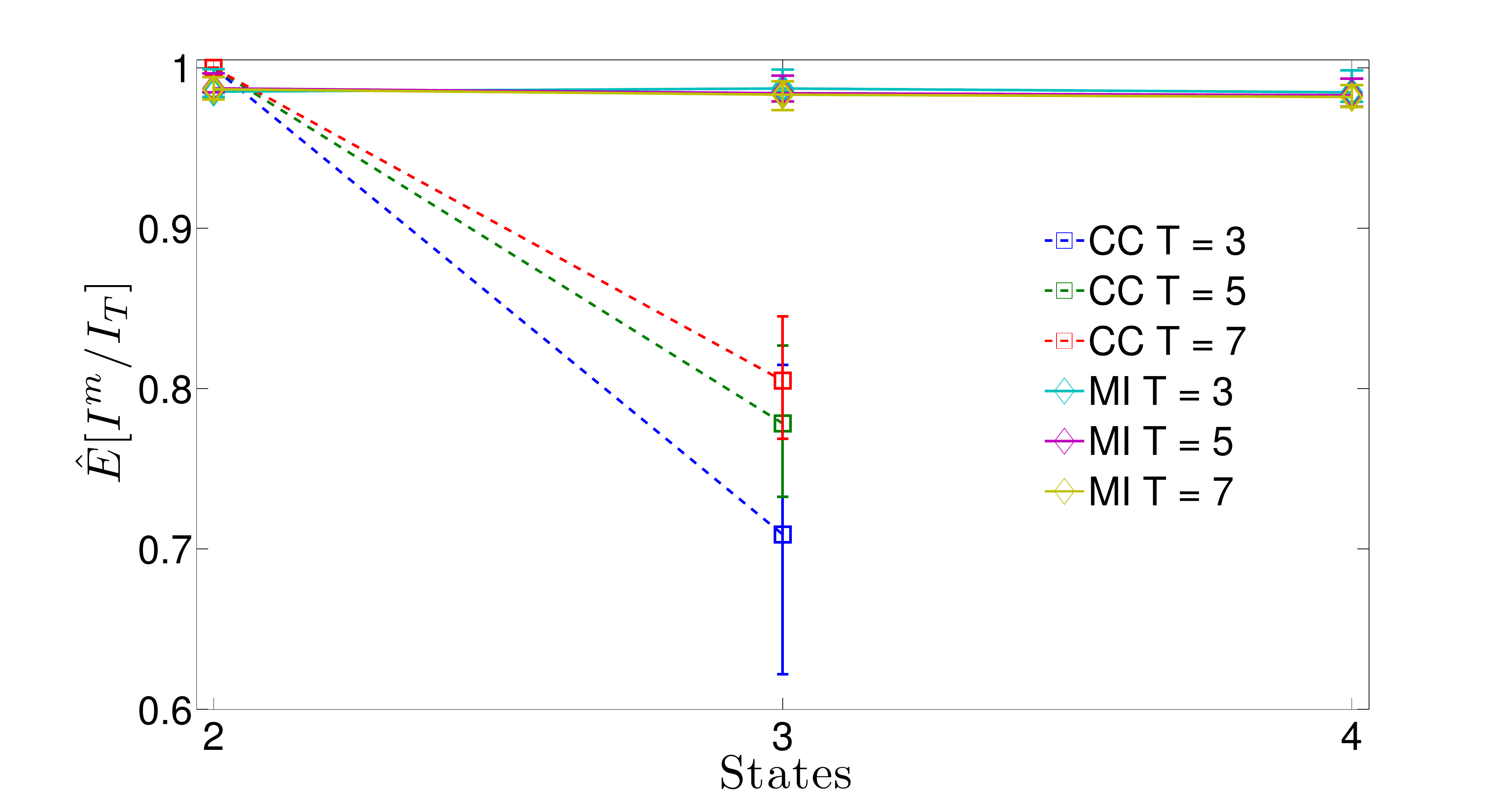}
   \end{center}
   \caption{\label{Fig:SubSampSynch} (Color online) The fraction of
     shared information coded by the mutual information (MI, diamonds
     with solid lines) and the cross-correlation (CC, squares with
     dashed lines) for a tuple of size T.  We simulated 100 nodes with
     $K = 4$, $G=0.001$, and the estimated expectations,
     $\hat{E}[...]$, are averages over 100 randomly selected tuples of
     the given size. All oscillators are synchronized, and their
     intrinsic frequencies are drawn from a normal distribution with
     zero mean and unit variance. Error bars are 25\% and 75\%
     quantiles.}
\end{figure}

\begin{figure}[!h]
  \begin{center}                         
    \includegraphics*[width=\columnwidth]{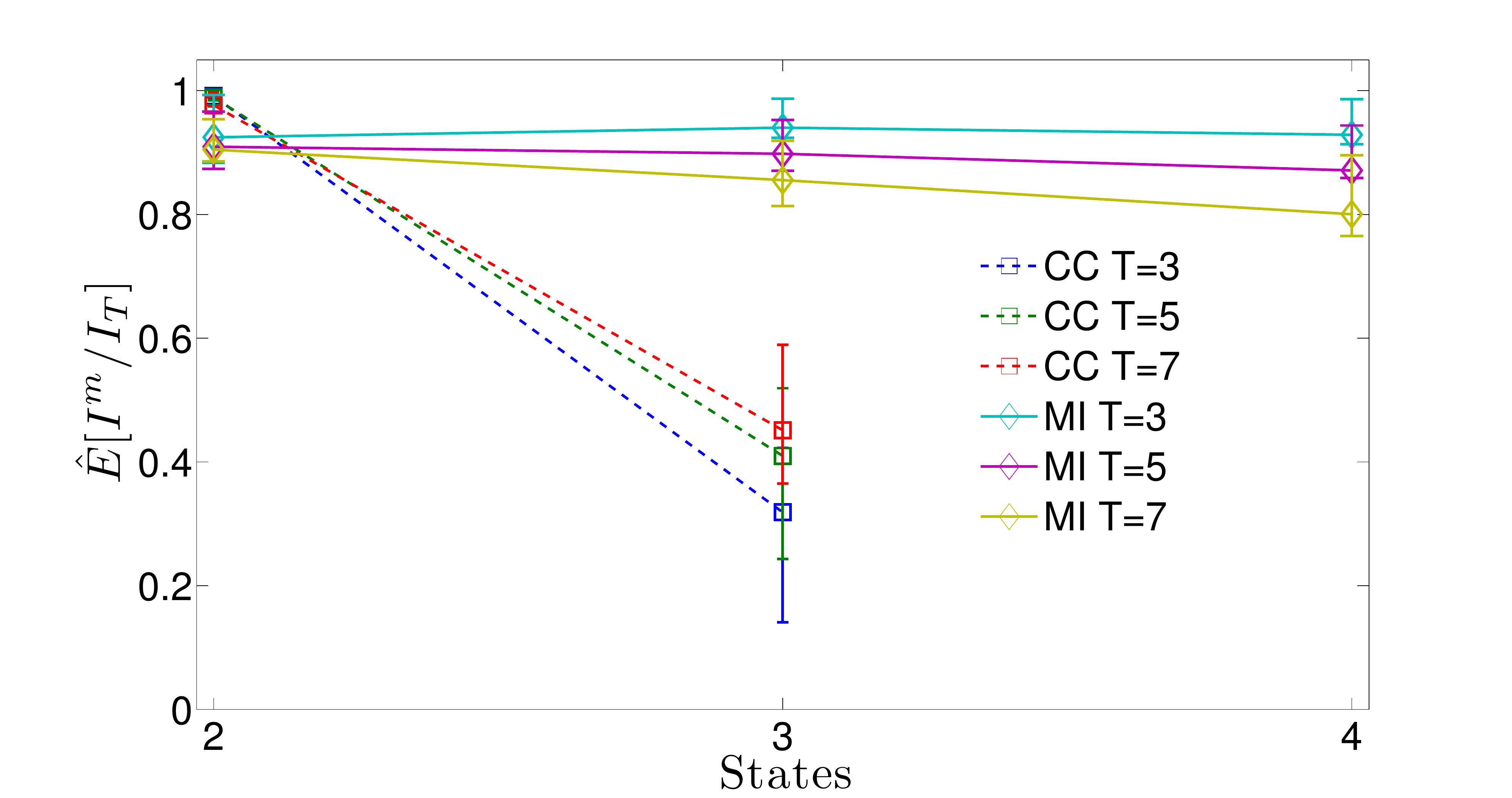}
   \end{center}
   \caption{\label{Fig:SubSampWeak} (Color online) The fraction of
     shared information coded by the mutual information (MI, diamonds
     with solid lines) and the cross-correlation (CC, squares with
     dashed lines) for a tuple of size T as in Fig.~\ref{Fig:SubSampSynch}
     but for weak coupling $K = 1.75$ leading to partial synchronization with more
     than 20 different clusters.}
\end{figure}

As in the analysis done in Ref.~\cite{Schneidman2006::Nat}, we
  analyzed the effects of sampling from a larger system by randomly
  selecting $T$ of the 100 oscillators and calculating $I^m/I_T$ for
  those oscillators. For each tuple size, $T$, we repeated this 100
  times, using the same sets of tuples in both regimes, and
  computed $\hat{E}[I^m/I_T]$ as the average of these values. As in
  our previous examples, our method outperforms the cross-correlation
  in the synchronized case (see Fig.~\ref{Fig:SubSampSynch}), as well
  as for weaker coupling (see Fig.~\ref{Fig:SubSampWeak}). Our method
  results in $\hat{E}[I^m/I_T] \lesssim 1$ in both regimes, and across
  all discretizations and tuple sizes, while the cross-correlation
  only does so for binary variables.

\paragraph{Resting-State Human Brain Networks:}

To illustrate the applicability of our methodology in real-world data
situations, we apply it to neuroimaging data, in a similar context as
in~\cite{Watanabe2013::NatCom}. In particular, we want to assess to
what extent the multivariate activity distribution is determined by
purely bivariate dependence patterns. The used data consist of time
series of functional magnetic resonance imaging signals from 96
healthy volunteers measured using a 3T Siemens Magnetom Trio scanner
in IKEM (Institute for Clinical and Experimental Medicine) in Prague,
Czech Republic. Average signals from 12 regions of the fronto-parietal
network were extracted using a brain atlas~\cite{Shirer2011}. After
preprocessing and denoising as in~\cite{Hlinka2011Neuroimage}, the
data were temporally concatenated. Each variable was further
discretized to 2 or 3 states using equiquantal binning.
Using our approach, we find $I^m/I_N=0.88$ for the 2-state and
$I^m/I_N=0.77$ for the
3-state discretizations, suggesting that bivariate dependence patterns
capture the dominant proportion of the information. For 2-state
discretization, this is smaller than
in~\cite{Watanabe2013::NatCom}. However, for the 3-state
discretization it provides a much higher estimate of the bivariate
dependence role than the method taking into account only correlations,
as in the case of the Kuramoto model.  This suggests that only when
accounting also for non-linear coupling, the bivariate dependencies
provide sufficient data structure approximation resolving the apparent
inconsistency of the results in~\cite{Watanabe2013::NatCom}. This is
also true for other brain networks~\cite{Martin2015::PRE}.

\paragraph{Discussion:}

Our method allows for potential speedups over the maximum entropy
calculation when conditioning on the bivariate distributions, as well
as when conditioning on the cross-correlations. In both of these cases
solving the associated Lagrange multiplier equations are non-linear
optimization problems. The maximum entropy distribution could also be
found using iterative fitting routines
like~\cite{Darroch1972::AnMatStat}, but in these cases the problem
will still scale like $n^N$ ($n$ is the cardinality of the
variables). While there are pathological linear optimization problems
that scale exponentially with $N$, there will always be a slightly
perturbed problem such that our method will scale
polynomially~\cite{Vershynin2009::SJC}.

Researchers have so far relied on conditioning on the
cross-correlations when insufficient data is available to estimate the
bivariate distributions. They either coarse grain to binary variables
where it is equivalent to conditioning on the
distributions~\cite{Watanabe2013::NatCom} --- potentially losing
important information --- or use higher cardinality variables where it
is only a linear approximation~\cite{Bialek2012::PNAS,Wood2012::PNAS}.
Our approach based on mutual information can be applied in these
cases; the associated entropies can be estimated with as few as
$2^{H/2}$ data points~\cite{Nemenman2011::Ent} ($H$ is measured in
bits). While this maximization has previously been prohibitively
difficult, our work shows that it is feasible allowing it to become
widely applicable and serve as a starting point before considering
multivariate information
measures~\cite{timme14a,Kralemann2014::NJP}. Additionally, if our
method returns a small $I^m/I_N$ this suggests both that the
faithfulness assumption used in causal inference is
violated~\cite{Eichler2013causal,Sun2014::Ent,Runge2012::PRL,Schindler2007::PR},
and that there is synergy among the variables~\cite{Griffith2014}.

Our calculation of $H_m$ for the mutual information is free of
distributional assumptions, computing the maximum entropy in the
general space of arbitrary cardinality variables. This may result in
higher entropy estimates than methods that consider predefined
cardinality, e.g., binary variables. Notably, our simulations suggest
that estimating $H_m$ in this way provides comparable, or
substantially lower, entropy estimates than $H_m$ for the
cross-correlation, which explicitly constrains the cardinality.  This
makes the technique competitive even when a specific cardinality could
be reasonably assumed.

\paragraph{Conclusions:}

In this work we introduced a novel method to determine the importance
of pairwise relationships by estimating the maximum entropy
conditioned on the mutual informations. We showed that by mapping this
problem to a linear optimization problem it could also be efficiently
computed.  Using the generic case of coupled oscillators we gave a
proof of principle example where our method was able to widely
out-preform conditioning on the cross-correlations. The example of the
resting-state brain network showed that this also carries over
to real world applications, highlighting the potential of the method
when cardinalities larger than two and nonlinear behavior are
important. 

Our results indicate that in many relevant cases functional
  networks based on mutual informations can in principle more
  accurately capture the dynamics of the system than those functional
  networks based on cross-correlations. These types of analyses should
  be applied \emph{before} studying functional networks, both to
  assess the validity of the network paradigm, as well as to test the
  appropriateness of using the given measure of association. Only high
  values in the fraction of shared information ensure that this is the
  case.  This has not been done in the vast majority of applications
  in the past. Due to the computational efficiency, our proposed
  methodology should allow to revisit this question, especially in
  areas where functional networks have already been widely applied
  such as in
  climate research~\cite{Donges2015::CD,Runge2015::NC,Martin2013::EPL, Deza2013}.


\begin{acknowledgments}
  This project was financially supported by NSERC (EM and JD) and by
  the Czech Science Foundation project No. 13-23940S and the Czech
  Health Research Council project NV15-29835A (JH). All authors would
  like to thank the MPIPKS for its hospitality and hosting the
  international seminar program ``Causality, Information Transfer and
  Dynamical Networks'', which stimulated some of the involved
  research. We also would like to thank P. Grassberger for many
  helpful discussions.
\end{acknowledgments}


%

\end{document}